\documentclass[aps,prl,showpacs,twocolumn,superscriptaddress]{revtex4}
\usepackage{graphicx}
\usepackage{bm}
\usepackage{amsmath,amssymb,amsfonts,bm}
\usepackage{epsfig}
\usepackage{epstopdf}
\usepackage{dcolumn}
\usepackage{grffile}
\usepackage{verbatim}
\usepackage{mathrsfs}

\usepackage[colorlinks=true,linkcolor=blue,citecolor=blue, urlcolor=blue]{hyperref}
\begin{document}


\title{Exact Solution to Interacting Kitaev Chain at Symmetric Point}

\author{Jian-Jian Miao}
\affiliation{Department of Physics, Zhejiang University, Hangzhou 310027, China}
\affiliation{Collaborative Innovation Center of Advanced Microstructures, Nanjing 210093,
China}

\author{Hui-Ke Jin}
\affiliation{Department of Physics, Zhejiang University, Hangzhou 310027, China}
\affiliation{Collaborative Innovation Center of Advanced Microstructures, Nanjing 210093,
China}

\author{Fu-Chun Zhang}
\affiliation{Kavli Institute for Theoretical Sciences, University of Chinese Academy of Sciences, Beijing 100190, China}
\affiliation{Department of Physics, Zhejiang University, Hangzhou 310027, China}
\affiliation{Collaborative Innovation Center of Advanced Microstructures, Nanjing 210093,
China}

\author{Yi Zhou}
\affiliation{Department of Physics, Zhejiang University, Hangzhou 310027, China}
\affiliation{Collaborative Innovation Center of Advanced Microstructures, Nanjing 210093,
China}

\date{\today}

\begin{abstract}

Kitaev chain model with nearest neighbor interaction U is solved exactly at the symmetry point $\Delta=t$ and chemical potential $\mu=0$ in open boundary condition. By applying two Jordan-Wigner transformations and a spin-rotation, such a symmetric interacting model is mapped to a non-interacting fermion model, which can be diagonalized exactly. The solutions include topologically non-trivial phase at $|U|<t$ and topologically trivial phase at $|U|>t$. The two phases are related by dualities. Quantum phase transitions in the model are studied with the help of the exact solution.

\end{abstract}

\maketitle


As a prototype of one-dimensional (1D) systems possessing Majorana zero modes (MZMs) \cite{Majorana} at two edges, Kitaev chain model\cite{Kitaev} has recently attracted a lot of attention \cite{Wilczek,Beenakker,Elliott}.
This non-interacting spinless fermion model was initially solved in a ring with periodic boundary condition.
Due to the increased interest on the effect of the interaction in MZMs \cite{Gangadharaiah,Stoudenmire,Thomale,Manolescu,Chan,Lutchyn,Sela,Cheng,Hassler,Klassen},
the model has been generalized to include nearest neighboring interaction. As examined by Fidkowski and Kitaev \cite{Fidkowski10}, such an interacting term may give rise to a transition from topological to  trivial phases in 1D, and the non-interacting classification of fermionic systems \cite{Ryu2008,Ryu2010,Kitaev2009} may ``collapse''.  The interaction Kitave chain has been studied by many authors \cite{Fidkowski11,Turner,Goldstein,Kells}, including by using numerical methods \cite{Rahmani1,Rahmani2,Milsted,Thomale2013,Iemini,Gergs,Hung,Miao2016}. On the other hand, the model does not have an analytic exact solution in the general case.  The exact ground states are available in a special set of tuned parameters\cite{Katsura}.  However,  the parameter space in their solvable model does not include any phase transition point.  In this Letter, we shall present an exact solution to the interacting Kitaev chain model at the symmetric point.  The solutions include the ground state and all excited states and show phase transitions from topologically non-trivial to trivial phases.

{\em Model:} We consider spinless fermions in a chain of length $L$ with open boundary condition. The Hamiltonian of such an interacting Kitaev chain reads
\begin{eqnarray}
H&=&\sum_{j=1}^{L-1}\left[-t\left(c_{j}^{\dagger}c_{j+1}+h.c.\right)+U\left(2n_{j}-1\right)\left(2n_{j+1}-1\right)\right.\nonumber\\
 &&\left.-\Delta\left(c_{j}^{\dagger}c_{j+1}^{\dagger}+h.c.\right) \right] -\mu \sum_{j=1}^{L}\left(n_{j}-{1\over2}\right),\label{eq:H}
\end{eqnarray}
where $c_{j}(c_{j}^{\dagger})$ is fermion annihilation (creation) operator on site $j$, $n_{j}=c_{j}^{\dagger}c_{j}$ is the fermion occupation number operator,
$t$ is the hopping integral, $\Delta$ is the $p$-wave superconducting pairing potential, $\mu$ is the chemical potential controlling the electron density, and $U$ is the nearest neighbor interaction.
Without loss of generality, both $t$ and $\Delta$ are chosen to be real and positive.
The parameter transformation of $\mu\to -\mu$ can be realized by the particle-hole conjugation $c_{j}\rightarrow\left(-1\right)^{j}c_{j}^{\dagger}$.  Therefore, $\mu=0$ corresponds to the particle-hole symmetry, which can be characterized by particle-hole conjugation operator $Z_{2}^{p}$ defined as follows.
\begin{equation}\label{def:Z2p}
Z_{2}^{p}=\prod_{j}\left[c_{j}+\left(-1\right)^{j}c_{j}^{\dagger}\right],
\end{equation}
$Z_{2}^p$ is conserved if and only if $\mu=0$. It is easy to verify that $(Z_{2}^p)^2=(-1)^L$ and $(Z_{2}^p)^{\dagger} Z_{2}^p=1$.
Hereafter, we shall assume that $L$ is an even number, so that $(Z_{2}^p)^2=1$ and $Z_{2}^{p}=\pm 1$.
Another good quantum number is the fermion number parity $Z_{2}^{f}$ defined as,
\begin{equation}\label{def:Z2f}
Z_{2}^{f}=e^{i\pi\sum_{j}n_{j}}=\left(-1\right)^{\hat{N}},
\end{equation}
where $\hat{N}=\sum_{j}n_{j}$ is the number of fermions in the system.  It is obvious that $(Z_{2}^f)^2=1$ and $[H,Z_{2}^f]=0$.
Both $Z_{2}^{p}$ and $Z_{2}^{f}$ will be used to characterize the ground states of the model Eq.~\eqref{eq:H} in different phases.

At $U=0$, the model is reduced to the usual non-interacting Kitaev chain model\cite{Kitaev}, which can be diagonalized exactly.
For interacting cases, $U\neq 0$, exact solution is not available in literature so far, except that the ground states have been constructed by Katsura {\it et al.}\cite{Katsura}
when chemical potential $\mu$ is tuned to a particular function of the other parameters $(t,\Delta,U)$.
In this Letter, we shall study the interacting Kitaev model at the symmetric point of $\Delta=t$ and $\mu=0$, and solve the model exactly by giving all the eigenstates.
Note that a similar symmetric model has been constructed in the context of Majorana linear chain without analytic solution\cite{Chiu}.

{\em Majorana fermion representation:} We shall study the Hamiltonian in Eq.~\eqref{eq:H} in the Majorana fermion representation.
Following Katsura {\it et al.}\cite{Katsura}, we split one complex fermion operator into two Majorana fermion operators $c_{j} = \frac{1}{2}\left(\lambda_{j}^{1}+i\lambda_{j}^{2}\right)$ and $c_{j}^{\dagger} = \frac{1}{2}\left(\lambda_{j}^{1}-i\lambda_{j}^{2}\right)$.
The Majorana fermion operators are real $\left(\lambda_{j}^{a}\right)^{\dagger}=\lambda_{j}^{a}$ and satisfy the anticommutation relations $\left\{ \lambda_{j}^{a},\lambda_{l}^{b}\right\} =2\delta_{ab}\delta_{jl}$,
where $a,b=1,2$. Thus, the Hamiltonian in Eq.~\eqref{eq:H} becomes
\begin{eqnarray}
H&=&\sum_{j=1}^{L-1} [-\frac{i}{2}\left(t+\Delta\right)\lambda_{j+1}^{1}\lambda_{j}^{2}-\frac{i}{2}\left(t-\Delta\right)\lambda_{j}^{1}\lambda_{j+1}^{2} \nonumber\\
&&-U\lambda_{j}^{1}\lambda_{j}^{2}\lambda_{j+1}^{1}\lambda_{j+1}^{2}] -\frac{i}{2}\mu \sum_{j=1}^{L}\lambda_{j}^{1}\lambda_{j}^{2}.\label{eq:HM}
\end{eqnarray}
At $U\neq 0$, the above Hamiltonian contains both quadratic and quartic terms, and can not be diagonalized straightforwardly.

{\em Mapping to a non-interacting chain:} The Hamiltonian in Eq.~\eqref{eq:HM} can be mapped to a non-interacting model consisting of quadratic terms only, at
$\Delta=t$ and $\mu=0$. The mapping is composed of two Jordan-Wigner transformations \cite{Jordan,Giamarchi} and a spin rotation.
Firstly, we construct spin operators by the first Jordan-Wigner transformation,
\begin{subequations}
\begin{eqnarray}\label{eq:JW1}
S_{j}^{x} & = & \frac{1}{2}\lambda_{j}^{1}e^{i\pi\sum_{l<j}n_{l}},\\
S_{j}^{y} & = & -\frac{1}{2}\lambda_{j}^{2}e^{i\pi\sum_{l<j}n_{l}},\\
S_{j}^{z} & = & \frac{i}{2}\lambda_{j}^{1}\lambda_{j}^{2}.
\end{eqnarray}
\end{subequations}
Thus, the Hamiltonian in Eq.~\eqref{eq:HM} can be written in terms of spin operators $S_{j}^{x}$ and $S_{j}^{z}$,
\begin{equation}\label{eq:HXZ}
H=\sum_{j=1}^{L-1}-4tS_{j}^{x}S_{j+1}^{x}+4US_{j}^{z}S_{j+1}^{z},
\end{equation}
which is a typical $XZ$ spin chain.

Secondly, we rotate all the spins by $\frac{\pi}{2}$ around the $x$-axis using the rotation operator $R=e^{-i\frac{\pi}{2}\sum_{j}S_{j}^{x}}$.
Then two new spin operators can be defined as $\tilde{S}_{j}^{x}:= R S_{j}^{x} R^{-1}={S}_{j}^{x}$ and $\tilde{S}_{j}^{y}:=R S_{j}^{y} R^{-1}={S}_{j}^{z}$.
The $XZ$ chain becomes a $XY$ chain,
\begin{equation}\label{eq:HXY}
H=\sum_{j=1}^{L-1}-4t\tilde{S}_{j}^{x}\tilde{S}_{j+1}^{x}+4U\tilde{S}_{j}^{y}\tilde{S}_{j+1}^{y}.
\end{equation}
Such an $XY$ spin chain has been exactly solved by Lieb, Schultz, and Mattis with the help of Jordan-Wigner transformation \cite{Lieb}.

Finally, following Lieb, Schultz, and Mattis, we use the second Jordan-Wigner transformation,
\begin{subequations}
\begin{eqnarray}\label{eq:JW2}
\tilde{S}_{j}^{x} & = & \frac{1}{2}\tilde{\lambda}_{j}^{1}e^{i\pi\sum_{l<j}\tilde{n}_{l}},\\
\tilde{S}_{j}^{y} & = & -\frac{1}{2}\tilde{\lambda}_{j}^{2}e^{i\pi\sum_{l<j}\tilde{n}_{l}},\\
\tilde{S}_{j}^{z} & = & \frac{i}{2}\tilde{\lambda}_{j}^{1}\tilde{\lambda}_{j}^{2},
\end{eqnarray}
\end{subequations}
to transform the $XY$ chain model in Eq.~\eqref{eq:HXY} to a quadratic fermion Hamiltonian, which is given by
\begin{equation}\label{eq:HR}
H=\frac{i}{2}\sum_{j,l=1}^{L}\tilde{\lambda}_{j}^{1}B_{jl}\tilde{\lambda}_{l}^{2},
\end{equation}
where $B_{jl}=2U\delta_{j,j+1}-2t\delta_{j,j-1}$ is a $L\times L$ real matrix. $\tilde{\lambda}_{j}^{1,2}$ can be written in terms of the original Majorana fermion operators $\lambda_{j}^{1,2}$ explictly,
\begin{subequations}\label{eq:JW12}
\begin{equation}
\tilde{\lambda}_{j}^{1}=\begin{cases}
\left(\prod_{l=odd}^{j-2}i\lambda_{l}^{2}\lambda_{l+1}^{1}\right)\lambda_{j}^{1}, & j=odd,\\
\left(\prod_{l=odd}^{j-3}i\lambda_{l}^{1}\lambda_{l+1}^{2}\right)i\lambda_{j-1}^{1}\lambda_{j}^{1}, & j=even,
\end{cases}
\end{equation}
and
\begin{equation}
\tilde{\lambda}_{j}^{2}=\begin{cases}
\left(\prod_{l=odd}^{j-2}i\lambda_{l}^{1}\lambda_{l+1}^{2}\right)i\lambda_{j}^{1}\lambda_{j}^{2}, & j=odd,\\
\left(\prod_{l=odd}^{j-3}i\lambda_{l}^{2}\lambda_{l+1}^{1}\right)\lambda_{j-1}^{2}i\lambda_{j}^{1}\lambda_{j}^{2}, & j=even.
\end{cases}
\end{equation}
\end{subequations}
With the help of Eqs.~\eqref{eq:JW12}, one is able to show that $\tilde{\lambda}_{j}^{1,2}$ are Majorana fermion operators by examining the relations
$\left(\tilde{\lambda}_{j}^{a}\right)^{\dagger}=\tilde{\lambda}_{j}^{a}$ and $\left\{\tilde{\lambda}_{j}^{a},\tilde{\lambda}_{l}^{b}\right\} =2\delta_{ab}\delta_{jl}$. 
So that the two sets of operators $\{\lambda_{j}^{a}\}$ and $\{\tilde{\lambda}_{j}^{a}\}$ must be related by a unitary transformation \cite{Kaufman}.

Thus, when $\Delta=t$ and $\mu=0$, the interacting fermion model given in Eq.~\eqref{eq:HM} with arbitary $U$ can be mapped to the non-interacting fermion model in Eq.~\eqref{eq:HR} through the unitary transformation given in Eq.~\eqref{eq:JW12}.
This is the central result in this Letter. 
Note that Gangadharaiah {\it et al.} \cite{Gangadharaiah} also pointed out that the interacting fermion model can be reduced to a free gapless fermion gas at the critical point $U=\Delta=t$ and $\mu=0$.

{\em Exact diagonalization:} The quadratic form of the Hamiltonian can be exactly diagonalized by singular value decomposition (SVD) as follows.
The non-symmetric matrix $B$ given in Eq.~\eqref{eq:HR} can be written in the SVD form $B=U\Lambda V^T$\cite{Katsura}, where $\Lambda$ is a non-negative diagonal matrix whose diagonal elements $\Lambda_k$ give rise to the singular values of $B$.
$U$ and $V$ are real orthogonal matrices and transform the Majorana fermion operators as
$\tilde{\lambda}_{k}^{1} = \sum_{j=1}^{L}U_{jk}\tilde{\lambda}_{j}^{1}$ and $\tilde{\lambda}_{k}^{2} = \sum_{j=1}^{L}V_{jk}\tilde{\lambda}_{j}^{2}$.
So that the self-conjugate and anticommutation relations remain the same,
$\left(\tilde{\lambda}_{k}^{a}\right)^{\dagger}=\tilde{\lambda}_{k}^{a}$ and $\left\{ \tilde{\lambda}_{k}^{a},\tilde{\lambda}_{q}^{b}\right\} =2\delta_{ab}\delta_{kq}$.

The diagonalized Hamiltonian reads
\begin{equation}\label{eq:HED}
H=\frac{i}{2}\sum_{k}\tilde{\lambda}_{k}^{1}\Lambda_{k}\tilde{\lambda}_{k}^{2}=\sum_{k}\Lambda_{k}\left(\tilde{c}_{k}^{\dagger}\tilde{c}_{k}-\frac{1}{2}\right),
\end{equation}
where $\tilde{c}_{k}=\frac{1}{2}\left(\tilde{\lambda}_{k}^{1}+i\tilde{\lambda}_{k}^{2}\right)$ and $\tilde{c}_{k}^{\dagger}=\frac{1}{2}\left(\tilde{\lambda}_{k}^{1}-i\tilde{\lambda}_{k}^{2}\right)$ are complex fermion operators.
Thus the energy spectrum is given by $\Lambda_{k}$, and reads
\begin{equation}
\Lambda_{k}=\sqrt{\left(t-U\right)^{2}\cos^{2}k+\left(t+U\right)^{2}\sin^{2}k},
\end{equation}
where each $k$-value gives rise to a single particle eigenstate in the rotated ($\tilde{\lambda}$) representation which will be called ``$k$-mode".
The values of $k$ and corresponding eigenstates can be determined similar to the non-interacting case \cite{Miao2016} (See Supplementary Materials for details).
The spectrum $\Lambda_{k}$ is gapful except at two quantum critical points $U=\pm t$.

There always exist $(L-1)$ real solutions and a single complex solution to $k$.
When $|U|>t$, the complex solution $k=k_{0}^{I}$ will give rise to corresponding singular value $\Lambda_{k_{0}^{I}}$, 
which is separated from the bulk energy continuum and has the asymptotic form at large $L$ as follows,
\begin{equation}\label{eq:Lk0I}
\Lambda_{k_{0}^{I}} \simeq \left(1-\left|\frac{t}{U}\right|\right)\left|\frac{t}{U}\right|^{L/2}.
\end{equation}
When $|U|<t$, the complex solution $k=k_{0}^{II}$ will give rise to corresponding singular value $\Lambda_{k_{0}^{II}}$, 
\begin{equation}\label{eq:Lk0II}
\Lambda_{k_{0}^{II}} \simeq \left(1-\left|\frac{U}{t}\right|\right)\left|\frac{U}{t}\right|^{L/2}.
\end{equation}

{\em Ground states:} In a finite system, the ground state $|0\rangle$ is non-degenerate and the energy spectrum is gapped except $U=\pm t$.
However, in the thermodynamic limit $L\to \infty$, $\Lambda_{k_0}\to 0$ exponentially, where $k_0=k_0^{I}$ ($k_0^{II}$) for $|U|>t$ ($|U|<t$). Thus, the first excited state $|1\rangle = c_{k_0}^{\dagger}|0\rangle$
is degenerate with the ground state $|0\rangle$ in the thermodynamic limit.

With the help of spin $XZ$ model given in Eq.~\eqref{eq:HXZ}, where long range spin correlation exists along $S^{x}$ or $S^{z}$ direction \cite{McCoy1968}, 
one is able to compute the long range density correlation, $\rho_{ij}=\left\langle(2n_{i}-1)(2n_{j}-1)\right\rangle$, in the bulk as follows,
\begin{equation}\label{eq:cor-density}
\lim_{|i-j|\rightarrow \infty }\rho_{ij} =\left\{ \begin{array}{cc}
\sqrt{1-(t/U)^{2}}, & U<-t, \\ 
0, & |U|<t, \\ 
(-1)^{i-j}\sqrt{1-(t/U)^{2}}, & U>t.
\end{array}
\right. 
\end{equation}
On the other hand, we have $\langle 0|n_{j}|0\rangle =\langle 1|n_{j}|1\rangle =\frac{1}{2}$. Therefore, the number fluctuation $\Delta N$ can be estimated for eigenstates $|0\rangle$ and $|1\rangle$: 
when $U<-t$, $\frac{(\Delta N)^{2}}{N^{2}}\rightarrow \sqrt{1-(t/U)^{2}}$; otherwise, $\frac{(\Delta N)^{2}}{N^{2}}\propto \frac{1}{N}$.
There are three different parameter regions for the symmetric interacting Kitaev chain model, $U<-t$, $-t<U<t$, and $U>t$.
(i) When $U<-t$, the ground state is a Shr\"{o}dinger-cat-like (CAT) state, which is a superposition of two trivial superconductor states with different occupation numbers,
$\langle\hat{N}\rangle/L\sim (1\pm\sqrt{1-(t/U)^{2}})/2$.\cite{comment1}
(ii) When $-t<U<t$, the ground state is a topological superconductor (TSC) state.
(iii) When $U>t$, the ground state is a charge density wave (CDW) state.

\begin{figure}[hptb]
\begin{center}
\includegraphics[width=8.4cm]{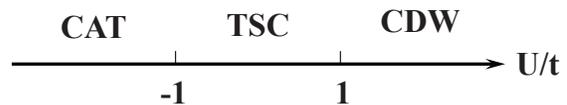}
\end{center}
\caption{Phase diagram of symmetric interacting Kitaev chain with $\Delta=t$ and $\mu=0$. CAT denotes a Shr\"{o}dinger cat-like state 
(a superposition of two trivial superconducting states with different occupation numbers), 
TSC denotes a topological superconducting state, and CDW denotes a charge density wave state.
Two critical points occur at $U=\pm t$.}
\label{fig:phase}
\end{figure}

For a finite system, the first excited state $|1\rangle$ can be distinguished from the ground state $|0\rangle$ by the fermion number parity $Z_{2}^{f}$ and the particle-hole conjugation $Z_{2}^{p}$,
when $|1\rangle$ is not degenerate with $|0\rangle$. Since $[Z_{2}^{f},H]=[Z_{2}^{p},H]=0$, any non-degenerate eigenstate $|\Psi_{n}\rangle$ of $H$ is also an eigenstate of $Z_{2}^{f}$ and $Z_{2}^{p}$,
namely, $Z_{2}^{f}|\Psi_{n}\rangle=\pm|\Psi_{n}\rangle$ and $Z_{2}^{p}|\Psi_{n}\rangle=\pm|\Psi_{n}\rangle$. With the help of the exact solution, one is able to show that
(1) when $|U|<t$, $\langle 1|Z_{2}^f|1\rangle = -\langle 0|Z_{2}^f|0\rangle$ and  $\langle 1|Z_{2}^p|1\rangle = \langle 0|Z_{2}^p|0\rangle$,
(2) when $|U|>t$, $\langle 1|Z_{2}^f|1\rangle = \langle 0|Z_{2}^f|0\rangle$ and  $\langle 1|Z_{2}^p|1\rangle = -\langle 0|Z_{2}^p|0\rangle$.
Now let $L\to \infty$ to approach the thermodynamic limit, these $Z_{2}^{f}$ and $Z_{2}^{p}$ values can be used to characterize different phases with two degenerate states $|0\rangle$ and $|1\rangle$.
For the TSC phase, $Z_2^f$ has opposite values ($\pm 1$) while $Z_2^p$ has the same value in the two degenerate ground states. For CDW and CAT phases, $Z_2^p$ has opposite values while $Z_2^f$ has the same value.

{\em Duality symmetries:}
There exist interesting dual relations between $|U|>t$ and $|U|<t$ phases when $\Delta=t$ and $\mu=0$, which impose quantum critical points at $U=\pm t$.
Such dualities can be seen clearly by rewriting Eq.~\eqref{eq:HR} as follows,
\begin{equation}\label{eq:HRD}
H=\frac{i}{2}\left(\sum_{j=1}^{L-1}2t\tilde{\lambda}_{j}^{2}\tilde{\lambda}_{j+1}^{1}+2U\tilde{\lambda}_{j}^{1}\tilde{\lambda}_{j+1}^{2}\right).
\end{equation}

For $U>0$, by interchanging the Majorana fermion operators in Eq.~\eqref{eq:HRD},
\begin{subequations}\label{eq:dual1}
\begin{equation}
\tilde{\lambda}_{j}^{1}\leftrightarrow\tilde{\lambda}_{j}^{2},
\end{equation}
Hamiltonian $H$ has the same form with parameters changes as
\begin{equation}
t\leftrightarrow U.
\end{equation}
\end{subequations}
Thus the duality between TSC and CDW phases has been established and $U=t$ must be the phase transition point separating these two phases.

For $U<0$, by interchanging the Majorana fermion operators as follows,
\begin{subequations}\label{eq:dual2}
\begin{equation}
\tilde{\lambda}_{j}^{1}\leftrightarrow(-1)^{j}\tilde{\lambda}_{j}^{2},
\end{equation}
$H$ will keep the same form with parameters changes as
\begin{equation}
t\leftrightarrow -U.
\end{equation}
\end{subequations}
Eqs.~\eqref{eq:dual2} set up the duality between TSC and CAT phases, and $U=-t$ must be the critical point separating these two phases.

It is interesting that the fermion number parity $Z_{2}^{f}$ and the particle-hole conjugation $Z_{2}^{p}$ will interchange to each other under the duality transformations.
In order to see this, we rewrite $Z_{2}^{f}$ and $Z_{2}^{p}$ in terms of $\tilde{\lambda}_{j}^{a}$,
\begin{subequations}\label{eq:Z2fp}
\begin{equation}
Z_{2}^{f}=\prod_{j=odd}^{L-1}(i\tilde{\lambda}_{j}^{1}\tilde{\lambda}_{j+1}^{2}),
\end{equation}
and
\begin{equation}\label{eq:Z2pR}
Z_{2}^{p} = \prod_{j=odd}^{L-1}(i\tilde{\lambda}_{j}^{2}\tilde{\lambda}_{j+1}^{1}).
\end{equation}
\end{subequations}
Therefore, we have the dual relation,
\begin{equation}
Z_{2}^{f} \leftrightarrow Z_{2}^{p}.
\end{equation}
It is noted that the fermion number parity in the rotated representation $\tilde{Z}_{2}^{f}$ is self-dual, say,
\begin{equation}
\tilde{Z}_{2}^{f} \leftrightarrow \tilde{Z}_{2}^{f}.
\end{equation}

{\em Phase transitions:}
The exact solution to the Hamiltonian $H$ also allows us to explore the quantum phase transition between neighboring phases.
The bulk spin correlation functions for the $XY$ spin chain have been evaluated by McCoy \cite{McCoy1968} as well as Capel and Perk \cite{Perk}.
In a previous paper, we proposed to use an edge correlation function to characterize the phase transition between TSC and SC phase\cite{Miao2016}, which is defined as follows,
\begin{equation}
G_{1L}=\left\langle i\lambda_{1}^{1}\lambda_{L}^{2}\right\rangle.
\end{equation}
In the rotated representation, it reads
\begin{equation}\label{eq:G1LR}
G_{1L}=\left\langle i\tilde{\lambda}_{1}^{1}\tilde{\lambda}_{L}^{2}\tilde{Z}_{2}^{f}Z_{2}^{f} \right\rangle.
\end{equation}
The detailed calculations for $G_{1L}$ can be found in the supplementary material.
For a generic ground state in the thermodynamic limit, the edge correlation function behaves as follows,
\begin{equation}
\lim_{L\rightarrow\infty}G_{1L}\propto \left\{ \begin{array}{cc} 1-\left(\frac{|U|}{t}\right)^{2}, & |U|<t,\\ 0, & |U|\geq t. \end{array} \right.
\end{equation}
The edge correlation function $G_{1L}$ is finite only in the TSC state in the thermodynamic limit.
Around the quantum phase transition points $U=\pm t$,
\begin{equation}
G_{1L}\propto (t-|U|)^z,
\end{equation}
with critical exponent $z=1$, which is the same as that for TSC to SC transition.

In summary, we study in this Letter interacting Kitaev chains with open boundary condition and $\Delta=t$ and $\mu=0$.
Exact solution has been obtained in the sense that all the eigenvalues and corresponding eigenstates are provided.
We find three different ground states which are used to define phases, a Shr\"{o}dinger-cat-like state for $U<-t$,
a topological superconductor state for $-t<U<t$, and a charge density wave state for $U>t$.
Duality symmetries between CAT and TSC and between TSC and CDW are found. Quantum phase transitions between neighboring phases are described by the edge correlation function,
and critical exponent is given as $z=1$. 
Besides the ground state properties discussed in this Letter, dynamics, thermodynamics and spectral function can be studied through the exact solution too. 
The interaction effect in the TSC phase can be observed by tunneling conductance $dI/dV\propto\rho(1,\omega)$ at edge. 
The edge density of states $\rho(1,\omega)$ is of the form $\rho(1,\omega)=A\delta(\omega)+B(\omega)\theta(\omega-\Delta_g)$,
where $\Delta_g$ is the bulk energy gap, $A\propto 1-(U/t)^2$ and $B(\omega)\propto \omega(\omega^2-\Delta_g^2)^{\alpha}$
with $\alpha\to \frac{1}{2}$ when $U\to 0$ and $\alpha\to -\frac{1}{2}$ when $|U|\to t$. The transition from positive to negative value of $\alpha$ exhibits 
qualitative difference in spectra and can be observed in tunneling experiments.

{\em Note added.} After our paper was posted on arXiv,  we received a note from Katsura, in which it was implied that the interacting fermion model can be reduced to a non-interacting fermion model at symmetric point \cite{Katsura2014}.

\emph{Acknowledgement.}
This work is supported in part by National Basic Research Program of China (No.2014CB921201/2014CB921203), National Key R\&D Program of the MOST of China (No.2016YFA0300202),
NSFC (No.11374256/11274269/11674278) and the Fundamental Research Funds for the Central Universities in China. F.C.Z was also supported by the University of Hong Kong's Grant Council via Grant No. AoE/P-04/08.

\section{Supplementary Materials}

\section{Self-conjugate and anticommutation relations for $\tilde{\lambda}_{j}^{a}$'s }\label{App:unitary}

Firstly, we shall prove the Majorana fermion operators $\tilde{\lambda}_{j}^{a}$ are real. Without loss of generality, we consider the Hermitian conjugate of $\tilde{\lambda}_{j}^{1}$ with odd $j$'s, and have
\begin{eqnarray}
\left(\tilde{\lambda}_{j}^{1}\right)^{\dagger} & = & \lambda_{j}^{1}\left(-i\lambda_{j-1}^{1}\lambda_{j-2}^{2}\right)\cdots\left(-i\lambda_{2}^{1}\lambda_{1}^{2}\right)\nonumber \\
 & = & \lambda_{j}^{1}\left(\lambda_{j-2}^{2}i\lambda_{j-1}^{1}\right)\cdots\left(\lambda_{1}^{2}i\lambda_{2}^{1}\right)\nonumber \\
 & = & \left(\lambda_{j-2}^{2}i\lambda_{j-1}^{1}\right)\cdots\left(\lambda_{1}^{2}i\lambda_{2}^{1}\right)\lambda_{j}^{1}\nonumber \\
 & = & \left(\lambda_{1}^{2}i\lambda_{2}^{1}\right)\cdots\left(\lambda_{j-2}^{2}i\lambda_{j-1}^{1}\right)\lambda_{j}^{1}\nonumber \\
 & = & \tilde{\lambda}_{j}^{1}.
\end{eqnarray}
So the Majorana fermion operator $\tilde{\lambda}_{j}^{1}$ with odd $j$'s is real. With the same procedure, we can show that $\tilde{\lambda}_{j}^{1}$ with even $j$'s and $\tilde{\lambda}_{j}^{2}$ are real too.

Secondly, we prove the Majorana fermion operators $\tilde{\lambda}_{j}^{a}$ satisfy the anticommutation relations $\left\{ \tilde{\lambda}_{j}^{a},\tilde{\lambda}_{l}^{b}\right\} =2\delta_{ab}\delta_{jl}$.
We first consider the two same Majorana fermion operators $\tilde{\lambda}_{j}^{a}$ at the same sites. Without loss of generality, we consider $\tilde{\lambda}_{j}^{1}$ with odd $j$'s at first,
\begin{eqnarray}
&   & \tilde{\lambda}_{j}^{1}\tilde{\lambda}_{j}^{1} \nonumber \\
& = & \left(\lambda_{1}^{2}i\lambda_{2}^{1}\right)\cdots\left(\lambda_{j-2}^{2}i\lambda_{j-1}^{1}\right)\lambda_{j}^{1}\left(\lambda_{1}^{2}i\lambda_{2}^{1}\right)\cdots\left(\lambda_{j-2}^{2}i\lambda_{j-1}^{1}\right)\lambda_{j}^{1}\nonumber \\
 & = & \left(\lambda_{1}^{2}i\lambda_{2}^{1}\right)^{2}\cdots\left(\lambda_{j-2}^{2}i\lambda_{j-1}^{1}\right)^{2}\left(\lambda_{j}^{1}\right)^{2}.
\end{eqnarray}
For $\left(\lambda_{l}^{a}\right)^{2}=1$, we have
\begin{eqnarray}
\left(\lambda_{l}^{2}i\lambda_{l+1}^{1}\right)^{2} & = & \left(-i\lambda_{l+1}^{1}\lambda_{l}^{2}\right)\left(\lambda_{l}^{2}i\lambda_{l+1}^{1}\right)\nonumber \\
 & = & -i\lambda_{l+1}^{1}i\lambda_{l+1}^{1}\nonumber \\
 & = & 1.
\end{eqnarray}
Therefore $\tilde{\lambda}_{j}^{1}\tilde{\lambda}_{j}^{1}=1$ when $j$ is odd. With the same procedure, we can show that $\tilde{\lambda}_{j}^{1}\tilde{\lambda}_{j}^{1}=1$ when $j$ is even and $\tilde{\lambda}_{j}^{2}\tilde{\lambda}_{j}^{2}=1$.
Then we consider the two Majorana fermion operators with the same index $a$ and at different sites $j$ and $l$. Without loss of generality, we consider $\tilde{\lambda}_{j}^{1}$ with odd $j$ and $j<l$. When $l$ is odd, we have
\begin{widetext}
\begin{eqnarray}
 \tilde{\lambda}_{j}^{1}\tilde{\lambda}_{l}^{1}+\tilde{\lambda}_{l}^{1}\tilde{\lambda}_{j}^{1}
 & = & \left[\left(\lambda_{1}^{2}i\lambda_{2}^{1}\right)\cdots\left(\lambda_{j-2}^{2}i\lambda_{j-1}^{1}\right)\lambda_{j}^{1}\right]\left[\left(\lambda_{1}^{2}i\lambda_{2}^{1}\right)\cdots\left(\lambda_{j}^{2}i\lambda_{j+1}^{1}\right)\cdots\left(\lambda_{l-2}^{2}i\lambda_{l-1}^{1}\right)\lambda_{l}^{1}\right]\nonumber \\
 &  & +\left[\left(\lambda_{1}^{2}i\lambda_{2}^{1}\right)\cdots\left(\lambda_{j}^{2}i\lambda_{j+1}^{1}\right)\cdots\left(\lambda_{l-2}^{2}i\lambda_{l-1}^{1}\right)\lambda_{l}^{1}\right]\left[\left(\lambda_{1}^{2}i\lambda_{2}^{1}\right)\cdots\left(\lambda_{j-2}^{2}i\lambda_{j-1}^{1}\right)\lambda_{j}^{1}\right]\nonumber \\
 & = & \left(\lambda_{1}^{2}i\lambda_{2}^{1}\right)^{2}\cdots\left(\lambda_{j-2}^{2}i\lambda_{j-1}^{1}\right)^{2}\left(\lambda_{j}^{2}i\lambda_{j+1}^{1}\right)\cdots\left(\lambda_{l-2}^{2}i\lambda_{l-1}^{1}\right)\lambda_{j}^{1}\lambda_{l}^{1}\nonumber \\
 &  & +\left(\lambda_{1}^{2}i\lambda_{2}^{1}\right)^{2}\cdots\left(\lambda_{j-2}^{2}i\lambda_{j-1}^{1}\right)^{2}\left(\lambda_{j}^{2}i\lambda_{j+1}^{1}\right)\cdots\left(\lambda_{l-2}^{2}i\lambda_{l-1}^{1}\right)\lambda_{l}^{1}\lambda_{j}^{1}\nonumber \\
 & = & \left(\lambda_{1}^{2}i\lambda_{2}^{1}\right)^{2}\cdots\left(\lambda_{j-2}^{2}i\lambda_{j-1}^{1}\right)^{2}\left(\lambda_{j}^{2}i\lambda_{j+1}^{1}\right)\cdots\left(\lambda_{l-2}^{2}i\lambda_{l-1}^{1}\right)\left\{ \lambda_{j}^{1},\lambda_{l}^{1}\right\} \nonumber \\
 & = & 0.
\end{eqnarray}
When $l$ is even, we have
\begin{eqnarray}
\tilde{\lambda}_{j}^{1}\tilde{\lambda}_{l}^{1}+\tilde{\lambda}_{l}^{1}\tilde{\lambda}_{j}^{1} & = & \left[\left(\lambda_{1}^{2}i\lambda_{2}^{1}\right)\cdots\left(\lambda_{j-2}^{2}i\lambda_{j-1}^{1}\right)\lambda_{j}^{1}\right]\left[\left(i\lambda_{1}^{1}\lambda_{2}^{2}\right)\cdots\left(i\lambda_{j}^{1}\lambda_{j+1}^{2}\right)\cdots\left(i\lambda_{l-3}^{1}\lambda_{l-2}^{2}\right)i\lambda_{l-1}^{1}\lambda_{l}^{1}\right]\nonumber \\
 &  & +\left[\left(i\lambda_{1}^{1}\lambda_{2}^{2}\right)\cdots\left(i\lambda_{j}^{1}\lambda_{j+1}^{2}\right)\cdots\left(i\lambda_{l-3}^{1}\lambda_{l-2}^{2}\right)i\lambda_{l-1}^{1}\lambda_{l}^{1}\right]\left[\left(\lambda_{1}^{2}i\lambda_{2}^{1}\right)\cdots\left(\lambda_{j-2}^{2}i\lambda_{j-1}^{1}\right)\lambda_{j}^{1}\right]\nonumber \\
 & = & \left[\left(\lambda_{1}^{2}i\lambda_{2}^{1}\right)\left(i\lambda_{1}^{1}\lambda_{2}^{2}\right)\cdots\left(\lambda_{j-2}^{2}i\lambda_{j-1}^{1}\right)\left(i\lambda_{j-2}^{1}\lambda_{j-1}^{2}\right)\right]\left(i\lambda_{j+1}^{2}\right)\left[\left(i\lambda_{j+2}^{1}\lambda_{j+3}^{2}\right)\cdots\left(i\lambda_{l-3}^{1}\lambda_{l-2}^{2}\right)i\lambda_{l-1}^{1}\lambda_{l}^{1}\right]\nonumber \\
 &  & +\left[\left(\lambda_{1}^{2}i\lambda_{2}^{1}\right)\left(i\lambda_{1}^{1}\lambda_{2}^{2}\right)\cdots\left(\lambda_{j-2}^{2}i\lambda_{j-1}^{1}\right)\left(i\lambda_{j-2}^{1}\lambda_{j-1}^{2}\right)\right]\left(-i\lambda_{j+1}^{2}\right)\left[\left(i\lambda_{j+2}^{1}\lambda_{j+3}^{2}\right)\cdots\left(i\lambda_{l-3}^{1}\lambda_{l-2}^{2}\right)i\lambda_{l-1}^{1}\lambda_{l}^{1}\right]\nonumber \\
 & = & 0.
\end{eqnarray}
\end{widetext}
By considering even $j$'s and $\tilde{\lambda}_{j}^{2}$, we have $\left\{ \tilde{\lambda}_{j}^{a},\tilde{\lambda}_{l}^{1}\right\} =0$ for $j\neq l$.
Finally we consider the two Majorana fermion operators with opposite indices $a=1$ and $a=2$. Without loss of generality, we begin with $\tilde{\lambda}_{j}^{1}$ when $j$ is odd and $j\leq l$. When $l$ is odd, we have
\begin{widetext}
\begin{eqnarray}
\tilde{\lambda}_{j}^{1}\tilde{\lambda}_{l}^{2}+\tilde{\lambda}_{l}^{2}\tilde{\lambda}_{j}^{1} & = & \left[\left(\lambda_{1}^{2}i\lambda_{2}^{1}\right)\cdots\left(\lambda_{j-2}^{2}i\lambda_{j-1}^{1}\right)\lambda_{j}^{1}\right]\left[\left(i\lambda_{1}^{1}\lambda_{2}^{2}\right)\cdots\left(i\lambda_{j}^{1}\lambda_{j+1}^{2}\right)\cdots\left(i\lambda_{l-2}^{1}\lambda_{l-1}^{2}\right)i\lambda_{l}^{1}\lambda_{l}^{2}\right]\nonumber \\
 &  & +\left[\left(i\lambda_{1}^{1}\lambda_{2}^{2}\right)\cdots\left(i\lambda_{j}^{1}\lambda_{j+1}^{2}\right)\cdots\left(i\lambda_{l-2}^{1}\lambda_{l-1}^{2}\right)i\lambda_{l}^{1}\lambda_{l}^{2}\right]\left[\left(\lambda_{1}^{2}i\lambda_{2}^{1}\right)\cdots\left(\lambda_{j-2}^{2}i\lambda_{j-1}^{1}\right)\lambda_{j}^{1}\right]\nonumber \\
 & = & \left[\left(\lambda_{1}^{2}i\lambda_{2}^{1}\right)\left(i\lambda_{1}^{1}\lambda_{2}^{2}\right)\cdots\left(\lambda_{j-2}^{2}i\lambda_{j-1}^{1}\right)\left(i\lambda_{j-2}^{1}\lambda_{j-1}^{2}\right)\right]\left(i\lambda_{j+1}^{2}\right)\left[\left(i\lambda_{j+2}^{1}\lambda_{j+3}^{2}\right)\cdots\left(i\lambda_{l-2}^{1}\lambda_{l-1}^{2}\right)i\lambda_{l}^{1}\lambda_{l}^{2}\right]\nonumber \\
 &  & +\left[\left(\lambda_{1}^{2}i\lambda_{2}^{1}\right)\left(i\lambda_{1}^{1}\lambda_{2}^{2}\right)\cdots\left(\lambda_{j-2}^{2}i\lambda_{j-1}^{1}\right)\left(i\lambda_{j-2}^{1}\lambda_{j-1}^{2}\right)\right]\left(-i\lambda_{j+1}^{2}\right)\left[\left(i\lambda_{j+2}^{1}\lambda_{j+3}^{2}\right)\cdots\left(i\lambda_{l-2}^{1}\lambda_{l-1}^{2}\right)i\lambda_{l}^{1}\lambda_{l}^{2}\right]\nonumber \\
 & = & 0.
\end{eqnarray}
When $l$ is even, we have
\begin{eqnarray}
\tilde{\lambda}_{j}^{1}\tilde{\lambda}_{l}^{2}+\tilde{\lambda}_{l}^{2}\tilde{\lambda}_{j}^{1} & = & \left[\left(\lambda_{1}^{2}i\lambda_{2}^{1}\right)\cdots\left(\lambda_{j-2}^{2}i\lambda_{j-1}^{1}\right)\lambda_{j}^{1}\right]\left[\left(\lambda_{1}^{2}i\lambda_{2}^{1}\right)\cdots\left(\lambda_{j}^{2}i\lambda_{j+1}^{1}\right)\cdots\left(\lambda_{l-1}^{2}i\lambda_{l}^{1}\right)\lambda_{l}^{2}\right]\nonumber \\
 & + & \left[\left(\lambda_{1}^{2}i\lambda_{2}^{1}\right)\cdots\left(\lambda_{j}^{2}i\lambda_{j+1}^{1}\right)\cdots\left(\lambda_{l-1}^{2}i\lambda_{l}^{1}\right)\lambda_{l}^{2}\right]\left[\left(\lambda_{1}^{2}i\lambda_{2}^{1}\right)\cdots\left(\lambda_{j-2}^{2}i\lambda_{j-1}^{1}\right)\lambda_{j}^{1}\right]\nonumber \\
 & = & \left(\lambda_{1}^{2}i\lambda_{2}^{1}\right)^{2}\cdots\left(\lambda_{j-2}^{2}i\lambda_{j-1}^{1}\right)^{2}\left(\lambda_{j}^{2}i\lambda_{j+1}^{1}\right)\cdots\left(\lambda_{l-1}^{2}i\lambda_{l}^{1}\right)\lambda_{j}^{1}\lambda_{l}^{2}\nonumber \\
 & + & \left(\lambda_{1}^{2}i\lambda_{2}^{1}\right)^{2}\cdots\left(\lambda_{j-2}^{2}i\lambda_{j-1}^{1}\right)^{2}\left(\lambda_{j}^{2}i\lambda_{j+1}^{1}\right)\cdots\left(\lambda_{l-1}^{2}i\lambda_{l}^{1}\right)\lambda_{l}^{2}\lambda_{j}^{1}\nonumber \\
 & = & \left(\lambda_{1}^{2}i\lambda_{2}^{1}\right)^{2}\cdots\left(\lambda_{j-2}^{2}i\lambda_{j-1}^{1}\right)^{2}\left(\lambda_{j}^{2}i\lambda_{j+1}^{1}\right)\cdots\left(\lambda_{l-1}^{2}i\lambda_{l}^{1}\right)\left\{ \lambda_{j}^{1},\lambda_{l}^{2}\right\} \nonumber \\
 & = & 0.
\end{eqnarray}
Hence we obtain $\left\{ \tilde{\lambda}_{j}^{1},\tilde{\lambda}_{l}^{2}\right\} =0$. Collecting all the above results, we prove that the Majorana fermion operators $\tilde{\lambda}_{j}^{a}$ satisfy the anticommutation relations $\left\{ \tilde{\lambda}_{j}^{a},\tilde{\lambda}_{l}^{b}\right\} =2\delta_{ab}\delta_{jl}$.
\end{widetext}

\section{$Z_2^f$ and $Z_2^p$ in terms of $\tilde{\lambda}_{j}^{a}$}
We derive the $Z_2^f$ and $Z_2^p$ symmetry operators in the rotated representation. For $Z_{2}^{f}$ symmetry operator we have\begin{eqnarray}
Z_{2}^{f} & = & e^{i\pi\sum_{j}n_{j}}\nonumber \\
 & = & e^{i\pi\sum_{j}\left(S_{j}^{z}+\frac{1}{2}\right)}\nonumber \\
 & = & e^{i\pi\sum_{j}\left(\tilde{S}_{j}^{y}+\frac{1}{2}\right)}\nonumber \\
 & = & \prod_{j}\left(-2\tilde{S}_{j}^{y}\right)\nonumber \\
 & = & \prod_{j}\left(\tilde{\lambda}_{j}^{2}e^{i\pi\sum_{l<j}\tilde{n}_{l}}\right),
\end{eqnarray}
where in the second line we use the first Jordan-Wigner transformation
to write the fermion operators $n_{j}$ in terms of spin operators
$S_{j}^{z}+\frac{1}{2}$. In the third line we adopt the rotation
$R$ to transform $S_{j}^{z}$ into $-\tilde{S}_{j}^{y}$. In the
fourth line we recall the identity $\exp\left(i\theta\vec{n}\cdot\vec{\sigma}\right)=I\cos\theta+i\vec{n}\cdot\vec{\sigma}\sin\theta$.
In the last line we use the second Jordan-Wigner transformation to
transformation the spin operators $\tilde{S}_{j}^{y}$ to Majorana
fermion operators in rotated representation. Using the following relation
\begin{equation}
\left(\tilde{\lambda}_{j}^{2}e^{i\pi\sum_{l<j}\tilde{n}_{l}}\right)\left(\tilde{\lambda}_{j+1}^{2}e^{i\pi\sum_{l<j+1}\tilde{n}_{l}}\right)=i\tilde{\lambda}_{j}^{1}\tilde{\lambda}_{j+1}^{2},
\end{equation}
we obtain the $Z_{2}^{f}$ symmetry operator in the rotated representation
\begin{equation}
Z_{2}^{f}=\begin{cases}
\prod_{j=odd}\left(i\tilde{\lambda}_{j}^{1}\tilde{\lambda}_{j+1}^{2}\right) & L=even,\\
\tilde{\lambda}_{1}^{2}\prod_{j=even}\left(i\tilde{\lambda}_{j}^{1}\tilde{\lambda}_{j+1}^{2}\right) & L=odd.
\end{cases}
\end{equation}

For $Z_{2}^{p}$ symmetry operator we have
\begin{eqnarray}
Z_{2}^{p} & = & \prod_{j}\left[c_{j}+\left(-1\right)^{j}c_{j}^{\dagger}\right]\nonumber \\
 & = & \begin{cases}
\prod_{j=odd}\left(i\lambda_{j}^{2}\lambda_{j+1}^{1}\right) & L=even,\\
i\lambda_{1}^{2}\prod_{j=even}\left(i\lambda_{j}^{2}\lambda_{j+1}^{1}\right) & L=odd,
\end{cases}
\end{eqnarray}
where we use the Majorana fermion representation. With the help of
Jordan-Wigner transformation, we have the relation
\begin{equation}
e^{i\pi\sum_{j}S_{j}^{x}}=\prod_{j}\left(2iS_{j}^{x}\right)=\prod_{j}\left(i\lambda_{j}^{1}e^{i\pi\sum_{l<j}n_{l}}\right).
\end{equation}
Using the following relation
\begin{equation}
\left(i\lambda_{j}^{1}e^{i\pi\sum_{l<j}n_{l}}\right)\left(i\lambda_{j+1}^{1}e^{i\pi\sum_{l<j+1}n_{l}}\right)=i\lambda_{j}^{2}\lambda_{j+1}^{1},
\end{equation}
we derive the identity
\begin{equation}
e^{i\pi\sum_{j}S_{j}^{x}}=\begin{cases}
\prod_{j=odd}\left(i\lambda_{j}^{2}\lambda_{j+1}^{1}\right) & L=even,\\
i\lambda_{1}^{2}\prod_{j=even}\left(i\lambda_{j}^{2}\lambda_{j+1}^{1}\right) & L=odd.
\end{cases}
\end{equation}
Thus we obtain the $Z_{2}^{p}$ symmetry operator after first Jordan-Wigner
transformation
\begin{equation}
Z_{2}^{p}=e^{i\pi\sum_{j}S_{j}^{x}}.
\end{equation}
With the same procedure, we obtain the $Z_{2}^{p}$ symmetry operator
in the rotated representation after rotation $R$ and the second Jordan-Wigner
transformation
\begin{eqnarray}
Z_{2}^{p} & = & e^{i\pi\sum_{j}\tilde{S}_{j}^{x}}\nonumber \\
 & = & \begin{cases}
\prod_{j=odd}\left(i\tilde{\lambda}_{j}^{2}\tilde{\lambda}_{j+1}^{1}\right) & L=even,\\
i\tilde{\lambda}_{1}^{2}\prod_{j=even}\left(i\tilde{\lambda}_{j}^{2}\tilde{\lambda}_{j+1}^{1}\right) & L=odd.
\end{cases}
\end{eqnarray}

\section{Edge correlation function $G_{1L}$}\label{App:edge-to-edge}
We calculate the edge correlation function $G_{1L}$ for the Hamiltonian $H$.
In the rotated representation, the edge correlation function $G_{1L}$ can be written as
\begin{eqnarray}
G_{1L} & = & \left\langle i\lambda_{1}^{1}\lambda_{L}^{2} \right\rangle = -4 \left\langle S_{1}^{x} S_{L}^{y} e^{i\pi \sum_{j=1}^{L-1} (S_{l}^z+{1\over2})} \right\rangle  \nonumber \\
 & = & -4 \left\langle S_{1}^{x} S_{L}^{x} e^{i\pi \sum_{j=1}^{L} (S_{l}^z+{1\over2})} \right\rangle  \nonumber \\
 & = & -4\left\langle S_{1}^{x}S_{L}^{x}Z_{2}^{f} \right\rangle = -4\left\langle \tilde{S}_{1}^{x}\tilde{S}_{L}^{x}Z_{2}^{f} \right\rangle \nonumber \\
 & = & \left\langle i\tilde{\lambda}_{1}^{1}\tilde{\lambda}_{L}^{2}\tilde{Z}_{2}^{f}Z_{2}^{f} \right\rangle ,
\end{eqnarray}
As $\left[Z_{2}^{f},H\right]=0$, we have $Z_{2}^{f}=\pm 1$ in all the eigenstates of $H$. For the ground state we choose $Z_{2}^{f}\left|0\right\rangle =\left|0\right\rangle $.
Thus the edge correlation function for the ground state $\left|0\right\rangle$ is given by
\begin{equation}
G_{1L}=\left\langle 0\right|i\tilde{\lambda}_{1}^{1}\tilde{\lambda}_{L}^{2}\tilde{Z}_{2}^{f}\left|0\right\rangle .
\end{equation}
Using the anticommutation relations of fermion operators, we have
\begin{subequations}
\begin{eqnarray}
e^{i\pi\tilde{n}_{j}} & = & -i\tilde{\lambda}_{j}^{1}\tilde{\lambda}_{j}^{2}\\
\tilde{\lambda}_{j}^{1}e^{i\pi\tilde{n}_{j}} & = & -i\tilde{\lambda}_{j}^{2}\\
\tilde{\lambda}_{j}^{2}e^{i\pi\tilde{n}_{j}} & = & i\tilde{\lambda}_{j}^{1}
\end{eqnarray}
\end{subequations}
The edge correlation function becomes
\begin{equation}
G_{1L}=\left(-i\right)^{N-1}\left\langle 0\right|\tilde{\lambda}_{1}^{2}\tilde{\lambda}_{2}^{1}\tilde{\lambda}_{2}^{2}\cdots\tilde{\lambda}_{L-1}^{1}\tilde{\lambda}_{L-1}^{2}\tilde{\lambda}_{L}^{1}\left|0\right\rangle .
\end{equation}
We use the Wick theorem to evaluate this expectation value. All kinds of the contractions consist of
\begin{subequations}
\begin{equation}
\left\langle \tilde{\lambda}_{j}^{1}\tilde{\lambda}_{l}^{1}\right\rangle =\sum_{k}U_{jk}U_{kl}^{T}=\delta_{jl},
\end{equation}
\begin{equation}
\left\langle \tilde{\lambda}_{j}^{2}\tilde{\lambda}_{l}^{2}\right\rangle =\sum_{k}V_{jk}V_{kl}^{T}=\delta_{jl},
\end{equation}
and
\begin{equation}
\left\langle \tilde{\lambda}_{j}^{2}\tilde{\lambda}_{l}^{1}\right\rangle =-\left\langle \tilde{\lambda}_{l}^{1}\tilde{\lambda}_{j}^{2}\right\rangle =-i\sum_{k}V_{jk}U_{kl}^{T}\equiv iC_{jl}.
\end{equation}
\end{subequations}
Since $\left\langle \tilde{\lambda}_{j}^{1}\tilde{\lambda}_{j}^{1}\right\rangle $ and $\left\langle \tilde{\lambda}_{j}^{2}\tilde{\lambda}_{j}^{2}\right\rangle $ never occur,
only $\left\langle \tilde{\lambda}_{j}^{2}\tilde{\lambda}_{l}^{1}\right\rangle $ terms contribute to $G_{1L}$. We have
\begin{eqnarray}
G_{1L} & = & \sum_{P}\left(-1\right)^{p}C_{1,P\left(2\right)}\cdots C_{L-1,P\left(L\right)}\nonumber \\
 & = & \left|\begin{array}{cccc}
C_{1,2} & C_{1,3} & \cdots & C_{1,L}\\
\vdots & \ddots &  & \vdots\\
\\
C_{L-1,2} & \cdots &  & C_{L-1,L}
\end{array}\right|,
\end{eqnarray}
where $P$ denotes all the permutation of $\left\langle \tilde{\lambda}_{1}^{2}\tilde{\lambda}_{2}^{1}\right\rangle \cdots\left\langle \tilde{\lambda}_{L-1}^{2}\tilde{\lambda}_{L}^{1}\right\rangle $ and $p$ is the sign associated with a given permutation.
The edge correlation function $G_{1L}$ is given by the cofactor of $\det C$,
\begin{equation}
G_{1L}=\left(-1\right)^{L+1}\left(C^{-1}\right)_{1L}\det C.
\end{equation}
Since the matrix $C$ is orthogonal
\begin{equation}
CC^{T}=\left(-VU^{T}\right)\left(-UV^{T}\right)=1,
\end{equation}
we have $\det C=\pm1$. The actual sign is determined as follows
\begin{equation}
\det C=\det\left(-VU^{T}\right)=\left(-1\right)^{L}\det V\det U^{T}.
\end{equation}
Since we have
\begin{equation}
B^{T}=V\Lambda^{T}U^{T},
\end{equation}
the determinant reads
\begin{equation}
\det C=\left(-1\right)^{L}\det B^{T}/\det\Lambda^{T}.
\end{equation}
As $\Lambda_{k}>0$ for all $k$, we obtain
\begin{equation}
\det C=\left(-1\right)^{L}\det B^{T}/\left|\det B^{T}\right|.
\end{equation}
For the present case,
\begin{equation}
\det B^{T}=\left(4tU\right)^{L/2},
\end{equation}
we have $\det C = \left[\text{sgn}(U)\right]^{L/2}$.  Thus the edge correlation function is
\begin{equation}
G_{1L}=\left[\text{sgn}(U)\right]^{L/2} C_{L1}=\left[\text{sgn}(U)\right]^{L/2}\sum_{k}U_{1k}V_{Lk}.
\end{equation}
We note that only the second kind of modes contribute to this correlation function. For $|U|<t$, the edge correlation function is given by
\begin{eqnarray}
G_{1L} & = & \left[\text{sgn}(U)\right]^{L/2}\left(U_{1k_{0}^{II}}V_{Lk_{0}^{II}}+\sum_{k^{II}}U_{1k^{II}}V_{Lk^{II}}\right)\nonumber \\
 & = & \left[\text{sgn}(U)\right]^{L/2}\left(-A_{k_{0}^{II}}^{2}\sinh^{2} v L-\sum_{k^{II}}A_{k^{II}}^{2}\delta_{k^{II}}\sin^{2}k^{II}L\right)\nonumber \\
 & = & \left[\text{sgn}(U)\right]^{L/2}\left[-\left(1-\left(\frac{U}{t}\right)^{2}\right)+O\left(1/L\right)\right].
\end{eqnarray}
For $|U|>t$, the edge correlation function is given by
\begin{eqnarray}
G_{1L} & = & \left[\text{sgn}(U)\right]^{L/2} \sum_{k^{II}}U_{1k^{II}}V_{Lk^{II}}\nonumber \\
 & = & -\left[\text{sgn}(U)\right]^{L/2} \sum_{k^{II}}A_{k^{II}}^{2}\delta_{k^{II}}\sin^{2}k^{II}L\nonumber \\
 & = & O\left(1/L\right).
\end{eqnarray}
The edge correlation function in the topological degenerate state $\left|1\right\rangle=\tilde{c}_{k_{0}^{II}}^{\dagger}\left|0\right\rangle $ can be calculated in the similar manner and is given by
\begin{eqnarray}
G_{1L} & = & \left\langle 1\right| i\lambda_{1}^{1}\lambda_{L}^{2} \left|1\right\rangle \nonumber \\
 & = & \begin{cases}
\left[\text{sgn}(U)\right]^{L/2}\left(1-\left(\frac{U}{t}\right)^{2}\right)+O\left(1/L\right) & |U|\leq t,\\
O\left(1/L\right) & |U|>t.
\end{cases}
\end{eqnarray}

\section{Spectral function $\rho\left(1,\omega\right)$ in TSC phase}\label{App:spectral}
To make connection with experiment, we calculate the spectral function $\rho\left(1,\omega\right)$, i.e. the local single-particle density
of states at edges in topological superconductor phase. To be simple, we only consider the local single-particle density of states at the edge site $j=1$.
The spectral function $\rho\left(1,\omega\right)$ dictates the $dI/dV$ signal in transport measurements or STM tunneling, which is defined as follows
\begin{equation}
\rho\left(1,\omega\right)=\rm{Im}G^{r}\left(\omega+i\eta\right),
\end{equation}
where $\rm{Im}$ takes the imaginary part, $\omega$ is the frequency,
$\eta$ is the positive infinitesimal and $G^{r}$ is the retarded Green's function given by
\begin{equation}
G^{r}\left(\omega\right)=\int_{-\infty}^{\infty}e^{-i\omega t}\theta\left(t\right)\left\langle c_{1}\left(t\right)c_{1}^{\dagger}\left(0\right)\right\rangle.
\end{equation}
Here we take the average $\left\langle \cdots\right\rangle $ with respect to the ground state $|0\rangle$. In terms of Majorana fermions,
\begin{eqnarray}
 &  & \left\langle c_{1}\left(t\right)c_{1}^{\dagger}\left(0\right)\right\rangle \nonumber \\
 & = & \frac{1}{4}\left\langle \tilde{\lambda}_{1}^{1}\left(t\right)\tilde{\lambda}_{1}^{1}\left(0\right)+i\tilde{\lambda}_{1}^{1}\left(t\right)\tilde{\lambda}_{1}^{2}\left(t\right)i\tilde{\lambda}_{1}^{1}\left(0\right)\tilde{\lambda}_{1}^{2}\left(0\right)\right\rangle ,
\end{eqnarray}
and other terms vanish. Here the Wick theorem is applicable since the Hamiltonian is bilinear in the rotated representation.
It is straightforward to see that
\begin{eqnarray}
 &  & \left\langle \tilde{\lambda}_{1}^{1}\left(t\right)\tilde{\lambda}_{1}^{1}\left(0\right)\right\rangle \nonumber \\
 & = & \left\langle \sum_{k}U_{k1}\tilde{\lambda}_{k}^{1}\left(t\right)\sum_{k'}U_{k'1}\tilde{\lambda}_{k'}^{1}\left(0\right)\right\rangle \nonumber \\
 & = & \sum_{k}U_{k1}^{2}e^{i\Lambda_{k}t}.
\end{eqnarray}
Similiar calculation shows that
\begin{equation}
\left\langle i\tilde{\lambda}_{1}^{1}\left(t\right)\tilde{\lambda}_{1}^{2}\left(t\right)i\tilde{\lambda}_{1}^{1}\left(0\right)\tilde{\lambda}_{1}^{2}\left(0\right)\right\rangle =\sum_{kk'}U_{k1}^{2}e^{i\Lambda_{k}t}V_{k'1}^{2}e^{i\Lambda_{k'}t}.
\end{equation}
By the Fourier transformation and taking the imaginary part, we finally obtain
\begin{equation}\label{eq:rho11}
\rho\left(1,\omega\right)=\sum_{k}U_{k1}^{2}\delta\left(\omega-\Lambda_{k}\right)+\sum_{kk'}U_{k1}^{2}V_{k'1}^{2}\delta\left(\omega-\Lambda_{k}-\Lambda_{k'}\right).
\end{equation}
In the TSC phase with $|U|<t$, the energy spectra contains a zero mode $k_0^{II}$ and gapped continuum with energy gap $\Delta_g=t-|U|$ in the thermodynamic limit.
Thus, in the frequency region $0<\omega<2\Delta_g$,
$\rho\left(1,\omega\right)$ is of the form
\begin{equation}
\rho\left(1,\omega\right)=A\delta\left(\omega\right)+B(\omega)\theta\left(\omega-\Delta_{g}\right).
\end{equation}
In the remaining part of this section, we shall study how the weight $A$ and the function $B(\omega)$ change as the interaction $U$.

Firstly, it is easy to see that
\begin{equation}
A=U_{k_{0}^{II}1}^{2}=A_{k_{0}^{II}}^{2}\sinh^{2}v^{II}L.
\end{equation}
In the thermodynamic limit $L\to \infty$, we have
\begin{equation}
e^{2v^{II}}\simeq\frac{t}{\left|U\right|},
\end{equation}
and
\begin{equation}
A_{k_{0}^{II}}^{2}\simeq4\left(1-\left(\frac{U}{t}\right)^{2}\right)\left(\frac{\left|U\right|}{t}\right)^{L}.
\end{equation}
Therefore
\begin{equation}
A\propto1-\left(\frac{U}{t}\right)^{2}.
\end{equation}

Secondly, we consider the $B(\omega)$ term. Using the relations $U_{k^{I}j}V_{k^{I'}j}=U_{k^{II}j}V_{k^{II'}j}=0$ and after straightforward algebra, we obtain that
\begin{eqnarray}
B\left(\omega\right) & = & \left(1-\frac{U^2}{t^2}\right)\sum_{k^{I}}A_{k^{I}1}^{2}\sin^{2}k^{I}L\delta\left(\omega-\Lambda_{k^{I}}\right)\nonumber \\
 & + & \sum_{k^{II}\neq k_{0}^{II}}A_{k^{II}1}^{2}\sin^{2}k^{II}L\delta\left(\omega-\Lambda_{k^{II}}\right).
\end{eqnarray}
Now we compute $B(\omega)$ in two limits, small $U$ limit with $|U|\ll t$ and critical limit with $|U|\simeq t$.

In the small $U$ limit with $\left|U\right|\ll t$, for $k^{I}$ modes we have
\begin{equation}
k^{I}\left(L+2\right)\simeq n\pi,
\end{equation}
where $n=1,2,\cdots,L$. So the $k$-dependent term in the factor
$A_{k^{I}}$ becomes
\begin{equation}
 \frac{\sin2k^{I}\left(L+1\right)}{\sin2k^{I}} \simeq  \frac{\sin\frac{2n\pi\left(L+1\right)}{L+2}}{\sin\frac{2n\pi}{L+2}}\nonumber  =  -\frac{\sin\frac{2n\pi}{L+2}}{\sin\frac{2n\pi}{L+2}}=-1.
\end{equation}
The factors $A_{k^{I}}$ can be approximated as
\begin{equation}
A_{k^{I}}\simeq\frac{2}{\sqrt{L+2}}.
\end{equation}
Thus the summand becomes
\begin{eqnarray}
 &  & A_{k^{I}1}^{2}\sin^{2}k^{I}L\nonumber \\
 & \simeq & \frac{4}{L+2}\sin^{2}\frac{n\pi}{L+2}L  =  \frac{4}{L+2}\sin^{2}\frac{2n\pi}{L+2}\nonumber \\
 & = & \frac{4}{L+2}\sin^{2}2k
\end{eqnarray}
The summation over $k^{I}$ modes can be replaced by the integral,
\begin{eqnarray}
 &  & \sum_{k^{I}}A_{k^{I}1}^{2}\sin^{2}k^{I}L\delta\left(\omega-\Lambda_{k^{I}}\right)\nonumber \\
 & = & \frac{4}{L+2}\int dk\sin^{2}2k\delta\left(\omega-\Lambda_{k^{I}}\right)\nonumber \\
 & = & \frac{4}{L+2}\int d\Lambda_{k^{I}}\frac{\Lambda_{k^{I}}}{4\sqrt{tU}\sqrt{\Lambda_{k^{I}}^{2}-\Delta_{g}^{2}}}4\frac{\Lambda_{k^{I}}^{2}-\Delta_{g}^{2}}{4tU}\delta\left(\omega-\Lambda_{k^{I}}\right)\nonumber \\
 & \propto & \frac{\omega\left(\omega^{2}-\Delta_{g}^{2}\right)^{1/2}}{\left(tU\right)^{3/2}}\theta\left(\omega-\Delta_{g}\right).
\end{eqnarray}
For $k^{II}$ modes, we have
\begin{equation}
k^{II}L\simeq n\pi,
\end{equation}
where $n=1,2,\cdots,L-1$. The summand becomes
\begin{equation}
A_{k^{II}1}^{2}\sin^{2}k^{II}L\simeq A_{k^{II}1}^{2}\sin^{2}\frac{n\pi}{L}L=0.
\end{equation}
So the contribution from $k^{II}$ modes can be neglected. The frequency dependence of $B\left(\omega\right)$ reads
\begin{equation}
B\left(\omega\right)\propto\omega\left(\omega^{2}-\Delta_{g}^{2}\right)^{1/2}
\end{equation}
in the small $U$ limit.

In the critical limit with $U\simeq t$, or equivalently $\left|U-t\right|/t\ll1$. Both $k^{I}$ and $k^{II}$ modes satisfy the equation
\begin{equation}
\sin k\left(L+2\right)\simeq\sin kL,
\end{equation}
or equivalently
\begin{equation}
\sin k\cos k\left(L+1\right)\simeq0,
\end{equation}
which give rise to
\begin{equation}
k=n\pi\,\, \text{ or }\,\, k\left(L+1\right)=\left(n+\frac{1}{2}\right)\pi.
\end{equation}
As the summand contain $\sin kL$, only the latter $k$'s contribute to the coefficient $B\left(\omega\right)$. The $k$-dependent term in the factor $A_{k}$ reads
\begin{equation}
\frac{\sin2k\left(L+1\right)}{\sin2k} \simeq  \frac{\sin\left(2n+1\right)\pi}{\sin2k}=0.
\end{equation}
The factors $A_{k}$ can be approximated as
\begin{equation}
A_{k}\simeq\frac{2}{\sqrt{L+1}}.
\end{equation}
The summand becomes
\begin{eqnarray}
 &  & A_{k1}^{2}\sin^{2}kL\nonumber \\
 & \simeq & \frac{4}{L+1}\sin^{2}\frac{\left(n+\frac{1}{2}\right)\pi}{L+1}L  =  \frac{4}{L+1}\cos^{2}\frac{\left(n+\frac{1}{2}\right)\pi}{L+1}\nonumber \\
 & = & \frac{4}{L+1}\cos^{2}k.
\end{eqnarray}
Replacing the summation in coefficient $B\left(\omega\right)$ by integral, we have
\begin{eqnarray}
 &  & \sum_{k}A_{k1}^{2}\sin^{2}kL\delta\left(\omega-\Lambda_{k}\right)\nonumber \\
 & = & \frac{4}{L+1}\int dk\cos^{2}k\delta\left(\omega-\Lambda_{k^{I}}\right)\nonumber \\
 & \propto & \omega\left(\omega^{2}-\Delta_{g}^{2}\right)^{-1/2}\theta\left(\omega-\Delta_{g}\right).
\end{eqnarray}
Near the quantum critical point $|U|\simeq t$, the frequency dependence of $B\left(\omega\right)$ reads
\begin{equation}
B\left(\omega\right)\propto\omega\left(\omega^{2}-\Delta_{g}^{2}\right)^{-1/2}.
\end{equation}

Thus we conclude that
\begin{equation}
B\left(\omega\right)\propto\omega\left(\omega^{2}-\Delta_{g}^{2}\right)^{\alpha},
\end{equation}
with $\alpha=1/2$ when $U\rightarrow0$ and $\alpha=-1/2$ when $\left|U\right|\rightarrow t$.


\begin{thebibliography}{99}
\bibitem{comment1} Note that nonzero $\mu$ will split the two-fold degeneracy in the thermodynamic limit and suppress the number fluctuation $\Delta N$, resulting in a trivial superconductor state.
However, a mini CAT state can survive in a finite system as long as the splitting $|\mu|L$ is small enough.
\end{thebibliography}

\end{document}